\documentstyle[12pt]{article}
\begin{document}
\title{An approach to Mel'nikov theory\\
in Celestial Mechanics}
\author{{G. Cicogna}\thanks{Email: cicogna@difi.unipi.it} 
\ and {M. Santoprete}\thanks{Present Address: Dept. of Mathematics 
and Statistics, Victoria B.C., Canada;} 
\thanks{Email: msantopr@math.uvic.ca} 
\\ Dipartimento di Fisica, Universit\`a di Pisa\\
Via Buonarroti 2, Ed. B \\ I-56127, Pisa, Italy}
\date{}
\maketitle
\begin{abstract} 
Using a completely analytic procedure -- based on a suitable extension 
of a classical method -- we discuss an approach to the 
Poincar\'e-Mel'nikov theory, which can be conveniently applied also 
to the case of non-hyperbolic critical points, and even if the 
critical point is located at the infinity. In this paper, we
concentrate our attention on the latter case, and  precisely on 
problems described by Kepler-like potentials in one or two degrees 
of freedom, in the presence of general time-dependent perturbations.
We show that the appearance of chaos (possibly including Arnol'd 
diffusion) can be proved quite easily and in a direct way, without 
resorting to singular coordinate transformations, such as the McGehee 
or blowing-up transformations. 
Natural examples are provided by the classical Gyld\'en problem, 
originally proposed in celestial mechanics, but also of interest 
in different fields, and  by the general 3-body problem in 
classical mechanics.
\end{abstract}
\def \bs{\bigskip}

\bs\bs
\def \pn{\par\noindent}
\quad
PACS(1999): 05.45.Ac, 45.05.+x, 95.10.Ce   

\def\R{{\bf R}}
\def \d{{\rm d}}

\def \nhy {nonhyperbolic}
\def \pd{\partial}
\def \pn{\par\noindent}

\def \en{\eqno}

\def\.#1{\dot #1}

~ \vskip 1 truecm

\parskip=0pt

\def\ref#1{$^{#1}$}

\def \de{\delta}
\def \th{\theta}
\def \Th{\Theta}
\def \ep{\epsilon}
\def \a {\alpha}
\def \ga{\delta}
\def \be{\beta}
\def \b{\beta}
\def \W{{\cal W}}
\def \r#1{$^{#1}$}
\def \( {\big( }
\def \) {\big) }
\def \intR {\int_{-\infty}^{+\infty}}
\def \grad {\nabla}
\def \ov{\over}
\def \G {{\pd W\big(R(t),t+t_0\big)\ov{\pd r}}}
\def \Gs{{\pd W\big(R(s),s+t_0\big)\ov{\pd r}}}
\def \q {\quad}
\def \qq {\qquad}
\def \bar{\overline}
\def \1{{\bf I}}
\def \2{{\bf II}}

\def \Ref{\pn{\bf References}\bs\pn\parskip=5 pt\parindent=0 pt}
\baselineskip 0.58 cm

\def\.#1{\dot #1}

\vfill\eject 

~ \vskip .5 truecm

In a previous paper\r{1}, we discussed a completely analytic procedure -- 
based on a suitable extension of a classical method\r{2} -- to introduce 
the Poincar\'e-Melnikov theory\r{3-5} concerning the appearance of chaotic 
behaviour, which can be applied even to the case where the 
critical point is {\it not hyperbolic}. The case of nonhyperbolic 
stationary points has been already considered by several authors, 
although with quite different methods or in different contexts: we 
mention here Ref.s 6-11; some other references, more strictly related 
to our arguments, will be given in the following. In Ref. 1, we have also 
shown that our procedure may work even in some problems where the 
critical point is located {\it at the infinity} of the real line, as 
occurs in the case of the classical Sitnikov problem\r{12}.

In this paper, we want first of all to reconsider and refine this 
approach, concentrating our attention precisely on problems described 
by Kepler-like potentials, in the presence of time-dependent 
perturbations of very general form. An example is provided by the 
classical Gyld\'en problem originally proposed in celestial mechanics, 
but also of interest 
in different fields (see Ref. {13}). We shall then extend this 
procedure to problems with 2 degrees of freedom  
and in the presence of rotational symmetry, and show that the appearance 
of chaos (possibly including Arnol'd diffusion\r{14}) 
can be proved quite easily and in a direct way, without resorting 
to singular coordinate transformations, such as the McGehee 
transformations\r{15}. 

Natural examples are provided by the general planar 3-body problem in 
classical mechanics\r{16-18}.

Let us remark that, although our procedure is quite general, 
we shall consider here -- for the sake of definiteness and clarity -- 
only the  case of the Kepler potential $V(r)=1/r$: 
from the following discussion it will appear
completely clear that the method can be easily applied  -- with 
suitable minor adjustments -- to different classes of problems, for 
instance to problems where the Kepler potential is replaced by another 
``long range'' potential, as e.g. $V\sim 1/r^\b$.

\vfill\eject
\bs\bs
\pn
{\bf I. The Gyld\'en-like problems}

\bs 
We start considering a 2-degrees of freedom problem described by a 
Hamiltonian of the form
$$H={1\over 2}{\bf p}^2-{1\over r} + \ep W(r,t)\en(1)$$
where ${\bf p}\equiv(p_1,p_2)$, i.e a standard (planar) Kepler problem 
plus a smooth perturbation $\ep W(r,t)$. This problem can be 
easily reduced to a 2-dimensional dynamical system for the two variables 
$r,\ \.r$, and for this reason we shall treat first this case, not only 
for better illustrating our procedure, but also in view of some direct 
and interesting applications, which include the classical Gyld\'en 
problem\r{13}.

Consider the ``parabolic'' solution of the unperturbed ($\ep=0$)
problem (1), which plays here the role of the homoclinic solution 
corresponding to the critical point at the infinity, i.e. 
$r=\infty,\ \.r=0$: it satisfies the equations
$$\.r=\pm {\sqrt{2r^2-k^2}\over r}\qq\qq\.\th={k\over{r^2}}\en(2) $$
where $k\not=0$ is the (constant) angular momentum, and the sign $-$ 
(resp. $+$) holds for $t<0$ (resp. $t>0$). From (2) one gets
$$\pm t={k^2+r\over 3}\sqrt{2r-k^2} + {\rm const}\qq\q
\th=\pm 2\arctan{\sqrt{2r-k^2}\over k}+ {\rm const}\en(3)$$
Let us denote by
$$R=R(t) \qq {\rm and}\qq \Th=\Th(t)\en(4)$$
the expressions giving the dependence of $r$ and of $\th$ on the time $t$ 
which are obtained ``inverting'' equations (3) with the conditions 
$R(0)=r_{min}=k^2/2$ and $\Th(0)=\pi$ (let us emphasize that it is 
{\it not} 
necessary, for our purposes, to have the explicit form of these functions). 
It will be useful only to remark that $R(t)$ is an even function 
and $\Th(t)$ is a odd function of time. The choice $\Th(0)=\pi$ 
corresponds to select the solution describing the parabola with axis 
coinciding with the $x_1$ axis and going to infinity when $x_1\to+\infty$; 
the whole Hamiltonian (1), including the 
perturbation $W(r,t)$, is rotationally invariant, and therefore this 
choice is clearly not restrictive. We refer to the next  section, 
where we shall consider symmetry-breaking perturbations, for some 
comment on this aspect.

The first point is now to find the negatively and positively asymptotic 
sets to the critical point $r=\infty,\ \.r=0$ 
in the presence of the perturbation $\ep W(r,t)$; notice that
this point would in fact correspond to the critical point 
$x=y=0$  under the McGehee's  singular coordinate 
transformation\r{12,13,15}
$$ r={1\over {x^2}} \qq\qq \.r=y\qq\qq \d t={1\ov{x^3}}\ \d s\en(5)$$
but -- instead of using this transformation (which usually requires 
quite cumbersome calculations), we introduce a direct method similar to 
the classical one used in Ref. {2} (see also Ref. {1}). Precisely, 
we want to show that  some natural assumptions on the perturbation $W(r,t)$ 
may guarantee not only the existence of smooth solutions  approaching 
$r=\infty,\ \.r=0$ for $t\to \pm\infty$, 
playing in this context the role of 
stable and unstable manifolds of the critical point at the 
infinity, but also the possible presence of infinitely many intersections 
of these asymptotic sets on the Poincar\'e sections.

It is important to remark that the critical point at the infinity 
is clearly not a hyperbolic point, and therefore the standard 
results of perturbation theory valid for hyperbolic points cannot be 
applied 
here. In particular, in order to be granted that the perturbation will 
preserve the property of the point at the infinity of being a critical
equilibrium, we have to impose that the perturbation vanishes at the 
infinity. We shall give the precise rate of this vanishing in the 
following \big(see eq. (18) below\big).

Let us write problem (1) in the form 
$$\ddot r={k^2\over{r^3}}-{1\over{r^2}}-\ep\ {\pd W(r,t)\ov{\pd r}}\en(6)$$
and let us look for solutions $r(t)$ of the perturbed problem (6) ``near'' 
the family of the homoclinic orbits $R(t-t_0)$: we then put (see Ref. {2})
$$r=R(t-t_0)+z(t-t_0) \en(7)$$
Inserting into (6) (with the time shift $t-t_0\to t$), we obtain
$$\ddot z+\Big({3k^2\over{R^4}}-{2\ov{R^3}}\Big)z=G\big(z(t),t+t_0\big)
\en(8)$$
where the r.h.s. \big(which we will shortly denote by 
$G(t,t_0)$\big) is given by 
$$G(t,t_0)\equiv -\ep\ {\pd W\big(R(t),t+t_0\big)\ov{\pd r}}+{\rm higher\ 
order\ terms\ in\ } z(t) \en(9)$$ 
Consider the homogeneous equation obtained putting $\ep=0$ in (8): one 
solution is clearly $\.R(t)$, another independent solution $\psi(t)$ can 
be constructed with standard methods (see e.g. Ref. {19}) or by direct 
substitution: these two solutions have a different behaviour for 
$t\to\pm\infty$, precisely:
$$\.R(t) \sim  \ |t|^{-1/3}\ \to 0  \qq {\rm and} \qq
           \psi(t) \sim  \ |t|^{4/3}\to\infty\en(10)$$ 
As well known, the general solution $z(t)$ of the complete (nonhomogeneous)
equation (8) 
can be written in the following integral form, 
with $A,B$ arbitrary constants and $t_1$ arbitrarily fixed\r{2,19}:
$$z(t)=A\.R(t)+B\psi(t)-\.R(t)\int_{t_1}^t \psi(s)G(s,t_0)\ \d s +
\psi(t)\int_{t_1}^t \.R(s)G(s,t_0)\ \d s \en (11)$$
Let us now look for solutions $z^{(-)}(t)$ \big(and resp. $z^{(+)}(t)$\big) 
of (11)
with the property of being bounded for $t\to-\infty$ (resp. $t\to+\infty$); 
these will provide solutions 
$$r^{(\pm)}(t)=R +z^{(\pm)} $$
of (6) which belong, by definition, to the  unstable and stable manifolds 
of the point $r=~\infty,\ \.r=0$. From (11), we have then to require that
the two quantities 
$$\psi(t)\Big(B+\int_{t_1}^t \.R(s)G(s,t_0)\ \d s\Big) \qq {\rm and}\qq
\.R(t)\Big(A-\int_{t_1}^t\psi(s)G(s,t_0)\ \d s\Big) \en(12)$$
remain bounded as $t\to -\infty$ when looking for the $r^{(-)}$ solutions, 
and resp. as $t\to +\infty$ for the $r^{(+)}$ solutions. 

Consider now the linearization of the problem (11) 
around the solution $z(t)\equiv 0$:  this amounts in particular to 
deleting the higher-order terms in the expression of $G$ in (9). 
Taking also into account the different behaviour given in (10) of 
the two fundamental solutions $\.R(t)$ and $\psi(t)$, it is easy to see
that the above conditions on the quantities (12) are  simultaneously 
satisfied both at $t=-\infty$ and at $t=+\infty$  
if for some $t_0$ the following Mel'nikov-type condition 
$$\intR \.R(t)\G\equiv M(t_0) =0 \en(13)$$
is fulfilled, together with the additional one, i.e. that the quantity
$$ \.R(t)\int_{t_1}^t\psi(s)\Gs\ \d s \en (14)$$
is bounded as $t\to\pm\infty$. Once these conditions are satisfied, one can 
conclude that, as a consequence of the implicit-function theorem\r{2} 
(or also thanks to a suitable version of the 
Lyapunov-Schmidt procedure, see e.g. Ref. {20}), there exists a smooth 
and bounded solution of (8). 

Let us now discuss the two above conditions (13-14).

The first one is identical to the usual Mel'nikov condition obtained 
under the standard assumption that the critical point is hyperbolic\r{3-5}: 
eq.(13) can therefore be viewed as an extension of the classical Mel'nikov 
formula to the present ``degenerate'' case, in which the critical point 
is at the infinity.
Let us now assume that the perturbation $W(r,t)$ is a smooth function,
periodic in the time $t$, with arbitrary period $T$ (it is not restrictive
to assume $T=2\pi$) and zero mean-valued:
$$\int_0^{2\pi} W(r,t) \d t =0 \en(15)$$
Then one can consider its Fourier expansion
$$W(r,t)=\sum_{n=1}^\infty \big(A_n(r)\cos nt +  B_n(r)\sin nt\big) 
\en(16)$$
and \big(thanks to the parity of the function $R(t)$\big)
write down the Mel'nikov function $M(t_0)$ defined in (13) in the form:
$$M(t_0)=\sum_{n=1}^\infty \big(\a_n\cos nt_0 + \b_n\sin nt_0\big)\en(17)$$
where
$$\a_n=\intR \.R(t){\d A_n\big(R(t)\big)\ov{\d r}}\sin nt\ \d t\q ,\q 
\b_n=\intR \.R(t){\d B_n\big(R(t)\big)\ov{\d r}}\sin nt\ \d t \en(17')$$
From this expression, one can immediately conclude that 
the function $M(t_0)$, being a smooth 
periodic function with vanishing mean value, 
must certainly possess zeroes, thus fulfilling condition (13).

For what concerns the second condition, which requires the boundedness 
of (14) and which appears here to compensate the lack of the
``exponential dichotomy'' peculiar of the hyperbolic case, a simple 
estimate of the behaviour as $|t|\to \infty$ of the integral in (14), 
using (10) and (16), and recalling from (3) that $R(t)\sim|t|^{2/3}$ 
as $|t|\to\infty$, shows that it is 
sufficient to assume that the quantities $A_n(r),\ B_n(r)$ in 
the expansion (16) vanish as $r\to\infty$ according to
$$A_n(r)\sim{a_n\ov{r^\ga}}\qq ,\qq B_n(r)\sim{b_n\ov{r^\ga}}\qq{\rm 
with}\qq \ga > 1/2 \en(18)$$
in order to be granted that the above condition on (14) is satisfied.
It can be noticed that the same condition (18) would also guarantee that 
under the McGehee transformation\r{12,13,15} (5), the perturbation is not 
singular at $x=0$.

Changing now the point of view, and considering the Poincar\'e sections of 
the $r^{(-)}$ and $r^{(+)}$ solutions, the above arguments show that, 
once conditions (13-14) are satisfied, there occurs a crossing of the 
negatively and positively asymptotic sets on the Poincar\'e section. One 
usually imposes at this point that the zeroes of the Mel'nikov function (13)
are  simple zeroes, i.e. that
$${\pd M\over{\pd t_0}}\not= 0 \en(19)$$
which ensures the transversality of the intersections, 
recalling that the function $M(t_0)$ 
expresses the signed distance between the intersecting manifolds\r{4,5}.
Actually, it can 
be noted that it is not strictly necessary to impose this condition, 
indeed -- according to an interesting and useful
result by Burns and Weiss\r{21} -- it is sufficient that the crossing is
``topological'', i.e., roughly, that there is really a ``crossing'', 
from one side to the other. But in our case this is certainly satisfied, 
because  $M(t_0)$, being a smooth periodic and zero mean-valued function, 
must necessarily change sign (see also Ref. {22} for a careful 
discussion on non-transversal crossings).
Using then standard arguments, which are not based on 
hyperbolicity, thanks to the periodicity of the perturbation,
one immediately deduces\r{2,4,5,21} that there is an infinite sequence of 
intersections, leading to a situation similar to the usual 
chain of homoclinic intersections typical of the homoclinic chaos. 

The presence of such infinitely many intersections is clearly reminiscent 
of the chaotic behaviour expressed by the Birkhoff-Smale theorem in terms 
of the equivalence to a symbolic dynamics described by the Smale 
horseshoes. Actually, this theorem
cannot be directly used in the present context because its standard 
proof is intrinsically based on hyperbolicity properties\r{4,5}. 
However, several arguments can be invoked even in the present 
``degenerate'' situation. First of 
all, for the case of degenerate critical points at the infinity, 
we can refer to the classical arguments used in Ref. {12}, and reconsidered 
by many others (see e.g. Ref.s {13,16,23}). More specifically, see
Ref. {24}, where an
equivalence to a ``\nhy\ horseshoe'' has been proved, in which
the contracting and expanding actions are not exponential but 
``polynomial'' in time. Let us also notice, incidentally, that the presence 
of Smale horseshoes and of a positive topological entropy has been proved 
by means of a quite general geometrical or
``topological'' procedure\r{21} which holds, in the presence of 
area-preserving perturbations, even in cases of
nonhyperbolic equilibrium points (i.e. not only in the case of 
degenerate critical points at the infinity, see Ref.s {1,25}).
Alternatively, in the general situation, one may possibly 
resort to the method of ``blowing-up'', devised to 
investigate the properties of \nhy\ singularities by means of suitable 
changes of coordinates\r{26,27}. 

Finally, for what concerns the regularity of the solutions and of the 
asymptotic sets, see in particular Ref.s {23-25,28}.

Summarizing, we can state the following:
\smallskip
\pn
{\bf Proposition 1.} {\it Consider a Kepler-like problem as in (1), and 
assume 
that the perturbation $W(r,t)$ is a smooth time-periodic function with 
zero mean value (16). Assume that it vanishes with $r\to\infty$ in 
such a way that (18) are satisfied. Then there is a chaotic behaviour of 
the solution, induced by a chain of infinitely many intersections 
in the Poincar\'e section of the negatively and positively 
asymptotic sets of the critical point at the infinity. }

\bs
Let us conclude this section with the obvious remark that the case of the 
Gyld\'en problem, for which the perturbation is given by
$$W(r,t)={\mu(t)\ov r}\en(20)$$
where $\mu(t)$ is a periodic function, satisfies all the above assumption 
and therefore exhibits chaotic behaviour\r{13}. The above discussion 
then generalizes this result to a larger class of problems and under 
weaker assumptions.
\bs\bs
\pn
{\bf II. Problems with 2 degrees of freedom}.

\bs
We now consider the case of planar Kepler-like problems as in (1) but in 
the presence of perturbations of the more general form $W=W(r,\th,t)$
$$H={1\over 2}{\bf p}^2-{1\over r} + \ep W(r,\th,t)\equiv
H_0+\ep W(r,\th,t)\en(21)$$
In this case the reduction of the problem as performed
in sect.~\1 is no longer possible, and we have to handle the four variables 
$x_1,p_1,x_2,p_2$ (or $r,\ \.r,\ \th,\ \.\th$). 
The first point to be remarked is that the degeneracy of the 
critical point at the infinity, $r=\infty,\ \.r=0$, appears  now even 
worse than before, indeed we 
have here a ``continuous family of points at the infinity'', due to the 
arbitrarity of the angle $\th$; more precisely, the homoclinic 
manifold, i.e. the set of solutions of the unperturbed equation  which 
are doubly asymptotic to $r=\infty,\ \.r=0$, is given here,
for each fixed value $k\not=0$ of the angular momentum, by the 2-dimensional 
manifold described by the family of parabolas of equations
$R=R(t-t_0),\ \Th(t-t_0)+\th_0$, where $R(t),\ \Th(t)$ have been defined 
in sect.~\1 (see (3-4)), with arbitrary $t_0,\th_0$, 
or --~in cartesian coordinates $u\equiv(x_1,p_1,x_2,p_2)$~-- by
$$\chi\equiv\chi(\th_0,t-t_0)\equiv \en(22)$$
$$\matrix{\Big(R(t-t_0)\cos\big(\Th(t-t_0)+\th_0\big), 
\qq\qq\qq\qq\qq\qq\qq\qq\qq\cr
\.R(t-t_0)\cos\big(\Th(t-t_0)+\th_0\big)-R(t-t_0)\.\Th(t-t_0)
\sin\big(\Th(t-t_0)+\th_0\big), \cr
\ R(t-t_0)\sin\big(\Th(t-t_0)+\th_0\big),
\qq\qq\qq\qq\qq\qq\qq\qq\qq\cr
\.R(t-t_0)\sin\big(\Th(t-t_0\big)+\th_0)+R(t-t_0)\.\Th(t-t_0)
\cos\big(\Th(t-t_0)+\th_0\big)\Big)}$$
In order to find conditions ensuring the occurrence of intersections 
of stable and unstable manifolds for the perturbed problem, we follow a 
similar (suitably extended) procedure as in sect.~\1. We first look for 
smooth 
solutions near the homoclinic manifold \big(see sect.~\1; here, clearly, 
$z\equiv(z_1,z_2,z_3,z_4)$\big)
$$u=\chi(\th_0,t-t_0)+z(\th_0,t-t_0) \en(23)$$
of the problem, which we now write in the form
$$\.u=J\grad_u H\equiv F(u)+\ep J\grad_u W\en(24)$$
($J$ being the standard symplectic matrix). Linearizing the 
problem along an arbitrarily fixed solution $\chi(\th_0,t-t_0)$ in 
the family (22), 
we get the following equation for $z(t)$
$$\.z=A(\th_0,t)z+ G(\th_0,t,t_0) \en(25)$$
where 
$$A(\th_0,t)=\big(\grad_uF\big)\big(\chi(\th_0,t)\big) \en(26) $$
and
$$G(\th_0,t,t_0)=\ep J\big(\grad_uW\big)\big(R(t),\Th(t)+\th_0,t+t_0\big) 
\en(27)$$
All the solutions $z(t)$ of (25) are are given by (cf. Ref. {19}; when not 
essential, the dependence on $\th_0,t_0$ will be sometimes dropped, 
for notational simplicity)
$$z(t)=z_h(t) + \Phi(t)\int_{t_1}^t \Phi^{-1}(s) G(s) \d s \en(28) $$
where $z_h(t)$ is any solution of the homogeneous linear problem
$$\.z=A(\th_0,t)z \en(29)$$
and $\Phi$ is a fundamental matrix of solutions of (29). 
There are certainly two solutions of (29), which are bounded for any 
$t\in\R$ (and vanish for $t\to\pm\infty$), namely 
$${\pd\chi\ov{\pd t}} \qq\qq {\rm and} \qq\qq {\pd\chi\ov{\pd\th}} \en(30)$$
as one may easily verify (this also follows from general 
arguments\r{29,30}). As seen in sect.~\1, due to the degeneracy of the 
critical point, instead of the exponential dichotomy, 
typical of the standard hyperbolic case, we now get a power behaviour 
$|t|^\sigma$  of the solutions, but the general arguments for controlling 
the behaviour for large $|t|$ of the solutions $z(t)$ in (28) can still 
be used.
Precisely (cf. Ref. {29}), observing also that, for Hamiltonian problems, 
the matrix $\Phi^{-1}J$
is the transposed of a fundamental matrix of solutions of the same
problem (29), one deduces from (28) and (30) that there exist
bounded solutions of (25), both for 
$t\to+\infty$ and for $t\to-\infty$, if the two following conditions are 
satisfied (cf. Ref.s {29,30})
$$ \intR \Big({\pd\chi(\th_0,t)\ov{\pd t}}\ , \grad_u 
W\big(R(t),\Th(t)+\th_0,t+t_0\big)\Big) \d t \equiv M_1(\th_0,t_0) =0 
\en(31)$$
and
$$\intR \Big({\pd\chi(\th_0,t)\ov{\pd \th}}\ , \grad_u 
W\big(R(t),\Th(t)+\th_0,t+t_0\big)\Big) \d t \equiv M_2(\th_0,t_0) =0 
\en(32)$$
where $\Big(\ \cdot\ ,\ \cdot\ \Big)$ stands for the scalar product 
in $\R^4$. Proceeding just as in sect.~\1, we assume that the perturbation 
$W(r,\th,t)$ is a smooth time-periodic function, and we still assume 
to hold a condition analogous to (18) in order to to guarantee
the boundedness of $z(t)$ at $t=\pm\infty$, as already discussed. 
The above conditions (31-32) can be more conveniently rewritten in the 
following form, using (22),
$$M_1(\th_0,t_0)=  \en(33)$$
$$\intR \!\!
\Big[\.R(t) {\pd W\big(R(t),\Th(t)+\th_0,t+t_0\big)\ov{\pd r}} + 
\.\Th(t) {\pd W\big(R(t),\Th(t)+\th_0,t+t_0\big)\ov{\pd\th}} \Big]\d t 
=0$$

$$M_2(\th_0,t_0)=
\intR {\pd W\big(R(t),\Th(t)+\th_0,t+t_0\big)\ov{\pd \th}}=0 \en(34)$$
It can be significant to remark here that it is easy to verify that 
these  conditions are
identical to the Mel'nikov conditions for the appearance of homoclinic 
intersections given e.g. in Ref.s  {5,31} in the standard hyperbolic case 
and deduced by means of a different procedure, namely
$$\intR \!\! \Big(\grad_u H_0,J\grad_u W\Big)\big(\chi(\th_0,t),t+t_0\big)\ 
\d t = \intR \{H_0,W\}\big(\chi(\ldots)\big)\ \d t = 0\en(35)$$
and
$$\intR \Big(\grad_u K,J\grad_u W\Big)\big(\chi(\th_0,t),t+t_0\big)\ 
\d t = \intR \{K,W\}\big(\chi(\ldots)\big)\ \d t = 0\en(36)$$
where $H_0$ is the unperturbed Hamiltonian, and $K=x_2p_1-x_1p_2$ is the 
angular momentum (which is indeed a constant of the motion for $H_0$).
We can then say that, again and apart from the additional condition (18), 
just as in sect.~\1, our procedure provides an extension of these 
formulas to the \nhy\ degenerate case we are considering here.

Notice also  that, as a consequence of the vanishing of the perturbation 
$W$ at $t=\pm\infty$, the first Mel'nikov condition
(33) can be written in the simpler form
$$M_1(\th_0,t_0)=\intR {\pd W\big(R(t),\Th(t)+\th_0,t+t_0\big)\ov{\pd t}} 
\d t=0 \en(33')$$
where clearly the derivative $\pd/\pd t$ must be performed only with respect 
to the explicit time-dependence of $W$.

It is also clear that, when the perturbation is independent of $\th$, as 
in the cases considered in sect.~\1, one consistently gets that the first 
condition (33) becomes just (13), 
whereas the second one (34) is identically satisfied.

Let us now introduce the ``Mel'nikov potential'' $\W=\W(\th_0,t_0)$ 
(cf. Ref. {32}), corresponding to the perturbation $W(r,\th,t)$:
$${\cal W}(\th_0,t_0)=\intR W\big(R(t),\Th(t)+\th_0,t+t_0\big) \d t 
\en(37)$$
then one gets from this definition and from (33$'$-34)
$$M_1(\th_0,t_0)={\pd \W\ov{\pd t_0}}=0 \qq\qq 
M_2(\th_0,t_0)={\pd \W\ov{\pd \th_0}}=0 \en(38)$$
In other words, the two Mel'nikov conditions are equivalent to the
existence of  
stationary points  for the Mel'nikov potential $\W(\th_0,t_0)$. 
On the other hand, $\W$ is a smooth doubly-periodic 
function, and such a function certainly possesses  points $\bar{\th_0},
\bar{t_0}$ where the two partial derivatives in (38) vanish, and this 
implies that the two conditions (33$'$-34) are certainly satisfied.

Now, exactly the same arguments (and with analogous remarks) 
presented in sect.~\1 show that the vanishing of the Mel'nikov functions 
entails the presence of a complicated dynamics, 
produced by the chain of the infinitely many intersections  of the 
asymptotic sets: see, e.g., Ref.s {5,28} for a detailed description 
of this  ``multidimensional'' case. 

In this context, we can also consider the appearance of  Arnol'd 
diffusion\r{14,33}. Indeed, the integrals of the Poisson brackets 
appearing in the Mel'nikov condition (35-36) give precisely 
the total amount of the ``variations'' produced by the perturbation, 
from $t=-\infty$ to $t=+\infty$, to the quantities $H_0$ and $K$ 
along the homoclinic solution $\chi$.
Proceeding in a similar way as 
in Ref. {14}, we can now look for the intersections of the asymptotic sets 
corresponding respectively for $t\to -\infty$ and for $t\to +\infty$ to 
{\it different} values $k_1$ and $k_2$ of the angular momentum $K$. 
Observing that for both these solutions the energy $H_0$ is the same, 
$H_0=0$, then -- following a by now classical idea\r{14} -- the above 
Mel'nikov conditions must be replaced by conditions of the form
$$M_1(\th_0,t_0)=0 \qq\qq M_2(\th_0,t_0)+k_1-k_2=0 \en(39)$$
and, exactly as in Ref. {14}, the conclusion is that, for  
$|k_1-k_2|$ small 
enough, intersections occur even from homoclinic solutions 
corresponding to different values of $K$. This argument is then consistent
with the occurence of Arnol'd 
diffusion: actually, a complete argument would necessitate a 
consideration of the role played, in the present ``degenerate'' 
situation, by  the usual ``non-resonance''  conditions: 
anyway, we point out that, in the same situation, i.e. in the case 
of general 3-body problem, but using a different approach, 
the occurrence of Arnol'd diffusion has 
been discussed in great detail by Xia\r{17-18}.
On the other hand, it is also known that the phenomenon of Arnol'd 
diffusion requires a quite delicate treatment: see Ref. 34 for a careful 
and updated discussion on this point.

Let us remark incidentally that the introduction of the above Mel'nikov 
potential $\W$ would be in general impossible if the perturbation is not 
Hamiltonian, i.e. if the problem (24) has the form of a general dynamical 
system
$$ \.u=J\nabla_uH_0+\ep g(u,t)\equiv F(u)+\ep g(u,t)\en(40) $$
where the perturbing term $g(u,t)$ \big(or $g(r,\.r,\th,\.\th,t)$\big) 
is ``generic''. Then in this case the above arguments cannot be repeated, 
and in particular the two Mel'nikov conditions, which can now be written 
in the general form
$$M_1(\th_0,t_0)=\intR\Big(\nabla_u H_0,g\Big)\big(\chi(\th_0,t),t+t_0\big)
\ \d t=0\en(41)$$
$$M_2(\th_0,t_0)=\intR\Big(\nabla_u K,g\Big)\big(\chi(\th_0,t),t+t_0\big)
\ \d t=0\en(42)$$
give two ``unrelated'' restrictions on $\th_0,t_0$, and one then remains 
with the problem of discovering if there are or not some 
$\bar{\th_0},\bar{t_0}$ which satisfy simultaneously both these conditions.


We can then summarize the above discussion in the following form.
\smallskip
\pn
{\bf Proposition 2.} {\it Let us consider a perturbed Kepler problem with 
Hamiltonian (21)
where $W$ is a smooth, time-periodic function, vanishing at the infinity 
according to (18). Then, there is a chaotic behaviour (possibly giving 
rise also to Arnol'd diffusion), 
induced by an infinite sequence of intersections in 
the Poincar\'e section of the negatively and positively asymptotic sets 
of the critical point at the infinity. In the case where the perturbed 
problem 
has the form (40) with a non-Hamiltonian perturbation $g(u,t)$, the 
same result is true if there are some $\bar{\th_0},\bar{t_0}$ satisfying 
simultaneously the two conditions (41-42).}

\bs\bs
\vfill\eject
\pn
{\bf III. Some applications and final remarks}

\bs
We shortly consider here some applications of the discussion presented in 
sect.~\2.  

As a first particular case, assume that the perturbation $W(r,\th,t)$ is
of the form
$$W=W(r,\ \a\th+bt)\en(43)$$
where $\a,b$ are arbitrary constants (the constant $\a$ should be clearly 
an integer, and $b\not=0$), then the two conditions (33$'$-34) actually 
coincide; observing on the other hand that the Fourier expansion of such 
a $W$ is a series as in (16) in terms of the single variable $\a\th+bt$, 
then 
these two conditions take the same form as in (17), and the Mel'nikov 
function is  a smooth periodic function of $\a\th_0+bt_0$, 
leading thus directly to the same conclusions obtained above 
for what concerns the existence of zeroes, and of their properties as well 
(it is really not a restriction to assume that $W$ is 
zero mean-valued, cf. Ref. {13}). 

A specially important example of this situation is provided 
by  the restricted circular 3-body problem, in this case indeed the 
perturbation is given by\r{16}
$$W={1\ov r}-{\cos(\th-t)\ov r^2}-{1\ov{\sqrt{1+r^2+2r\cos(\th-t)}}}
\en(44)$$
and therefore just one condition has to be considered. The presence of 
the chaos produced by the chain of intersections of the asymptotic sets 
then is automatically granted by our discussion.
Notice that the above expression (44) is actually the first-order 
expansion of the full potential in terms of the parameter $\ep$ (which in 
this case is given by the mass ratio $\mu$ between two celestial bodies), 
but also the exact expression of this potential, as given in Ref. {16}, is 
in fact a function of $\th-t$ only.

It is completely clear that, in the presence of a more 
general perturbation, e.g. of the  form (just to give an example)
$$W(r,\th,t)=W_{(1)}(r,\th-t)+W_{(2)}(r,\th+t)$$
one now obtains two different Mel'nikov conditions, and the presence of a 
chaotic behaviour follows from the existence of simultaneous solutions 
$\bar{\th_0},\bar{t_0}$, which is ensured by the arguments shown in sect.~\2.

The above results hold essentially unchanged if the perturbation depends 
on two (or more) parameters $\ep_1,\ep_2,\ldots$, i.e. if one assumes 
that $W$ may be written in the form (cf. Ref. {29})
$$W=\ep_1W_{(1)}(r,\th,t)+\ep_2W_{(2)}(r,\th,t)+\ldots \en(45)$$
A natural example is provided by the planar 3-body problem, where one has 
to deal, in the more general elliptic case, with a quite complicated 
expression of the perturbation containing three parameters 
(different masses and eccentricity\r{17-18}). Let us remark, however, 
that, at least in the simpler case of restricted elliptic problem, in 
which one has two parameters 
($\ep_1=\mu$ is one mass ratio and $\ep_2=e$ the eccentricity),
the perturbation cannot be written as in the above ``first-order'' 
form (45), but rather it takes the form\r{17}
$$W=\mu\big(W_{(1)}(r,\th,t)+e\ W_{(2)}(r,\th,t)\big)\en(46)$$
in which the eccentricity plays the role of a ``second-order'' 
perturbation.
On the basis of our previous arguments, we can just say that the presence of 
zeroes of the Mel'nikov functions, as obtained above for the circular case 
$\ep_2=e=0$, cannot be destroyed  by the higher-order perturbation 
due to the eccentricity, and therefore chaos should be expected to 
persist in the elliptic case, whereas Arnol'd diffusion should appear as
a second-order effect.
A complete study of the general 3-body problem, and a full discussion of 
its chaotic properties, including the appearance of Arnol'd diffusion, 
together with several other dynamical features, is given in Ref.s 16-18.

\vfill\eject

\Ref
\r{1}  G. Cicogna, M. Santoprete, 
Phys. Letters A {\bf 256}, 25-30 (1999)

\ref{2}  S.-N. Chow, J.K. Hale and J. Mallet-Paret,  J. Diff. Eq.
{\bf 37}, 351-373 (1980) 

\ref{3}  V.K. Mel'nikov, 
Trans. Moscow Math. Soc. {\bf 12},  1-56 (1963)

\ref{4}   J. Guckenheimer and P.J. Holmes, {\it Nonlinear oscillations, 
dynamical 
systems and bifurcations of vector fields} (Springer, Berlin 1983)

\ref{5}  S. Wiggins, {\it Global bifurcations and chaos} (Springer, Berlin 
1989)

\r{6} S. Schecter, SIAM J. Math. Anal. {\bf 18}, 1142-1156 and 
1699-1715 (1987) 

\r{7} S. Schecter, Nonlinearity {\bf 3}, 79-99 (1990)

\r{8} B. Deng, SIAM J. Math. Anal. {\bf 21}, 693-720 (1990)
          
\r{9} F. Battelli, Ann. Mat. Pura Appl. {\bf 166}, 267-289 (1994) 

\r{10} F. Battelli,  Boll. Un. Mat. Ital. B {\bf 8}, 87-110 (1994)

\r{11} J.H. Sun and D.J. Luo, Sci. China A {\bf 37}, 523-534 (1994)

\ref{12}  J. Moser, {\it Stable and random motions in dynamical systems} 
(Princeton Univ. Press, Princeton 1973)

\ref{13}  F.  Diacu  and  D. Selaru, 
     J.   Math.  Phys.  {\bf  39}, 6537-6546 (1998)

\r{14} V.I. Arnol'd, Dokl. Akad. Nauk. SSSR {\bf 156},  9-13 (1964)

\r{15}  R. McGehee,  J. Diff. Eq. {\bf 14}, 70-88 (1973), see also Invent. 
Math. {\bf 27}, 191-227 (1974) 

\ref{16} Z. Xia, 
J. Diff. Eq. {\bf 96}, 170-184 (1992)

\r{17} Z. Xia, J. Dynamics Diff. Eq. {\bf 5}, 219-240 (1993)

\ref{18} Z. Xia, 
J. Diff. Eq. {\bf 110}, 289-321 (1994)

\ref{19}  E.A. Coddington  and  N. Levinson, {\it Theory of ordinary 
differential equations} (McGraw-Hill, New York 1955)

\r{20} J.K. Hale, in {\it Bifurcation theory and applications} (L. 
Salvadori ed.), (Springer, Berlin 1984), p.106-151

\r{21}  K. Burns and H. Weiss, 
{\it  Comm. Math. Phys.} {\bf 172}, 95-118 (1995), see also 
A. J. Homburg and H. Weiss, 
preprint 1999, Maryland and Pennsylvania St. Universities  

\r{22}  V. Rayskin, preprint 1998, Dept. of Math., Texas Univ., Austin

\ref{23}  Xiao-Biao Lin,  Dynamics Reported {\bf 5}, 99-189 (1996)

\r{24}  H. Dankowicz and P. Holmes,  J. Diff. Eq. {\bf 116}, 468-483 (1995)

\ref{25}   J. Casasayas, E. Fontich and  A. Nunes, Nonlinearity {\bf 
5}, 1193-1210 (1992), and Nonlinear Diff. Eq. Appl. {\bf 4}, 201-216 (1997) 
 
\ref{26}  F. Dumortier,  J. Diff. Eq. {\bf 23}, 53-106 (1977)

\ref{27}  M. Brunella and M. Miari,  J. Diff. Eq. {\bf 85}, 338-366 (1990) 

\r{28}  C. Robinson, J. Diff. Eq. {\bf 52}, 356-377 (1984) 

\r{29} J. Gr\"undler, SIAM J. Math. Anal. {\bf 16}, 907-931 (1985)

\ref{30}   M. Santoprete, Thesis, Dept. of Physics, Univ. of Pisa, 1999

\r{31} C. Robinson,  Contemp. Math. {\bf 198}, 45-53 (1996) 

\r{32} A. Delshams and R. Ram{\'\i}rez-Ros,  
Comm. Math. Phys. {\bf 190}, 213-245 (1997) 

\r{33} P.J. Holmes and J.E. Marsden,  J.  Math.  Phys.  {\bf  23}, 
669-675 (1982)

\r{34} P. Lochak, in  {\it Hamiltonian systems with 
three or more degrees of freedom}, (NATO ASI Proceedings, C. Simo ed.), 
(Kluwer, Boston, Mass.  1999), p. 168-183

\end{document}